# Multi-task Driver Steering Behaviour Modeling Using Time-Series Transformer

Yang Xing, *Member, IEEE,* Wenbo Li, Xiaoyu Mo, Chen Lv, *Senior Member, IEEE*

*Abstract*—Human intention prediction provides an augmented solution for the design of assistants and collaboration between the human driver and intelligent vehicles. In this study, a multi-task sequential learning framework is developed to predict future steering torques and steering postures based on the upper limb neuromuscular Electromyography (EMG) signals. A single-right-hand driving mode is particularly studied. For this driving mode, three different driving postures are also evaluated. Then, a multi-task time-series transformer network (MTS-Trans) is developed to predict the steering torques and driving postures. To evaluate the multi-task learning performance, four different frameworks are assessed. Twenty-one participants are involved in the driving simulator-based experiment. The proposed model achieved accurate prediction results on the future steering torque prediction and driving postures recognition for single-hand driving modes. The proposed system can contribute to the development of advanced driver steering assistant systems and ensure mutual understanding between human drivers and intelligent vehicles.

## I. INTRODUCTION

Intelligent vehicles are showing their tremendous potential in the improvement of traffic safety, efficiency, and diversity [1,2]. Although a series of achievements have been obtained in the past decade, a challenging question still needs to be answered, which is how to efficiently design collaboration and interaction mechanisms for a human driver and intelligent vehicles [3]. Before fully autonomous vehicles can be realized, human drivers will still need to share the vehicle control authority with automation. In this situation, the mutual understanding in terms of anticipating the teammates' physiological and psychological states such as the intention, attention, and behaviours will enable these two intelligent agents to interact efficiently with each other [4]. The lateral steering collaboration between the human and intelligent vehicles or partially automated driving vehicles (ADVs) can contribute to a safer steering control under normal or critical situations [5]. Understanding human steering patterns and predicting driver steering intention will benefit the driver-in-the-loop advanced driver assistance system (ADAS) and shared steering control system [6]. Meanwhile, the risk assessment and hazard prediction system can be developed based on the driver steering intention to prevent the human driver from making dangerous driving manoeuvres [7]. Besides, as the driver will share the control authority with the ADVs, it is essential to estimates the driver's steering intention and quality before the driver can take back the control authority. Such a function could be especially important when the driver has a lower level of situation awareness dues to the recovery from the secondary tasks [8].

Considering these superiorities, in this study, a deep time-series modelling approach for sequential steering torque and driving posture prediction is proposed. The predicted steering torque can optimize the design of advanced shared steering control algorithms, intention-aware take-over systems, etc. Recognition of steering posture can also help to understand the steering patterns and steering qualities so that accurate assistance can be provided [9]. Besides, the recognition of specific driving postures can also help to analyse driving/steering behaviours, fatigue, and effectiveness [6]. This study particularly focuses on a single-hand driving mode. The reasons are compared to both-hand studies, the single-hand driving mode is less studied [10], and the steering pattern can be more complex than the both-hand driving mode [11]. For the single-hand driving mode, the 3-clock, 130-clock, and 12-clock driving postures are investigated. Last, four MTS-Trans frameworks are evaluated to analyse the multi-task learning and inference performance.

In sum, the contribution of this study can be summarized as: first, a multi-task time-series framework for driver steering intention prediction with respect to different driving postures is developed. The framework can build a connection between the driver's neuromuscular dynamics and future steering torque within a certain horizon. Second, the impact of different driving postures is studied based on the proposed model to exploit how different driving behaviours influence driver intention prediction. Last, quantitative analysis and comparison for the MTS-Trans frameworks are proposed considering different driving modes, postures, and features.

## II. LITERATURE REVIEW

A large number of existing studies on driver steering intention focus on the modelling of steering behaviours using driver models, driver states, and vehicle dynamics information [12,13]. The vision-based approaches have been widely studied in the past for discrete steering intention prediction, which may contain lane change intention inference and turn manoeuvre prediction [14]. In [15], a Bernoulli heatmap approach was developed for a convolution neural network for driver's head pose estimation. In [16], a calibration-free eye

Y. Xing is with the Centre of Autonomous and Cyber-Physical Systems, Cranfield University, Bedfordshire, MK43 0AL, U.K. (e-mail: yang.x@cranfield.ac.uk).

Wenbo Li is with the School of Vehicle and Mobility, Tsinghua University, Beijing, 100084, China (e-mail: wenboli@tsinghua.edu.cn).

Y. Mo, is with the School of Mechanical and Aerospace Engineering, Nanyang Technological University, 639798, Singapore. (e-mail: xiaoyu006@ntu.edu.sg).

C. Lv, is with the School of Mechanical and Aerospace Engineering, Nanyang Technological University, 639798, Singapore. (e-mail: lyuchen@ntu.edu.sg).

gaze estimation approach was developed for driver attention understanding. Despite vision-based methods, researchers have developed several successful steering intention classification systems based on different sensor systems. For example, the brain-machine interface for the human driver steering intention prediction using the electroencephalogram (EEG) signals were developed in [17]. Meanwhile, the vehicle dynamic information from the CAN bus, such as the steering force, vehicle heading, and lateral acceleration, can also be used for accurate steering intention recognition. For instance, the lane change manoeuvre of the driver can be recognized based on an explicit mathematical model of the steering behaviour using vehicle dynamic states [18]. However, using vehicle dynamics-based information is difficult to make an in-advance prediction for the steering intention as the steering intention can only be reflected by the vehicle motion after the driver has initiated the steering manoeuvre. In general, driver intention can be recognized with features from the in-cabin vision system, vehicle states, human physical and cognitive states (heart rate and EEG), etc. For instance, a vision-based system can capture driver behaviours such as head pose, driving behaviours, and intent [19]. However, the prediction of continuous steering intention and steering torque would pose a significant challenge to the vision-based systems dues to the high diversity of steering behaviours, postures, and habits for each driver.

Although the neuromuscular dynamics for the steering tasks have achieved a series of successful studies, most of the existing studies lack consideration of the shared steering with the autonomy counterpart, such as the ADVs. Meanwhile, to enhance long-term temporal feature representation, the Transformer network has been developed since 2017 [20]. Then, it was found that a Transformer-based network can achieve superior results in many areas such as natural language processing, image processing, and time-series modelling. Compared to the popular time-series methods like autoregressive integrated moving average (ARMAX) [21], recurrent neural network (RNN), and long-short term memory (LSTM) [22], etc, transformer-based methods have a significant advantage over many time-series modelling tasks [23, 24]. To enable efficient shared steering between the human and vehicle, a mutual understanding and intent communication scheme will be needed. Thus, a novel continuous steering intent prediction model considering different driving modes and driving postures is developed in this study to contribute to the mutual understanding between the driver and the intelligent vehicle.

### III. Experiment Design

In this section, an experiment design for the simulator-based steering intention prediction system is proposed. Then, signal collection and processing for the EMG data are described.

#### A. Experiment Platform and Testing Scenarios

The experiment testbed was developed on a six-degrees-of-freedom driving simulator. The testbed can be used for a wide range of human-in-the-loop steering experiments [6]. The CarSim system was used to develop the simulation environment. A DynPick WEF-6A1000 force sensor and TR-60TC torque angle sensor, which were implemented under the steering wheel of the driving simulator, were used to collect the real-time steering dynamics. The Nihon Kohden ZB-150H wireless sensors were used for the collection of EMG signals with a sampling frequency of 1000 Hz.

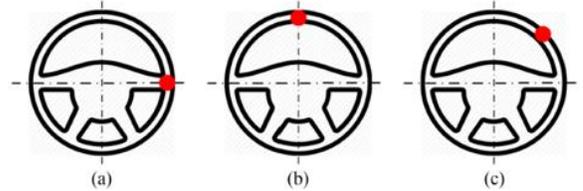

Figure 1. Illustration of the three kinds of driving postures for the single-hand driving mode.

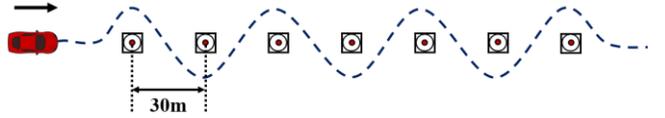

Figure 2. The simulated Slalom driving task used in this study.

TABLE 1. SUMMARY OF THE DETECTED NEUROMUSCULAR SIGNALS FOR THE SINGLE-HAND DRIVING MODES

| Driving Mode | Detected Neuromuscular Signals |
| --- | --- |
| Single-hand (right) | PMA-C, DELT-A, DELT-P, TM, TB-L, deltoid middle (lateral) (DELT-M), triceps lateral head exterior (TB-LAT), biceps (BC), infraspinatus (INFT), pectoralis major (PM) |

During the experiment, the participants were required to drive the simulator with the single-hand driving mode using three different driving postures (as shown in Figure 1). Ten EMG electrodes are used to extract the upper limb neuromuscular dynamics, respectively. The detailed neuromuscular signals are further summarized in Table 1. Twenty-one voluntary male participants were involved in the experiment. All of the participants were deeply informed about the purpose and risks of this experiment and agreed to participate. The 21 participants were divided into three groups, namely, skilled group, average group, and unskilled group, according to their reported driving experience. The introduction of different driving experience improves the pattern diversity of the experiment data and help to avoid bias patterns in the analysis of driver neuromuscular dynamics.

All the participants were required to control the steering wheel by following a constant sinusoidal input. The participants were required to perform a slalom driving task during the experiment with around 60 km/h velocity (as shown in Figure 2), which is an international standard testing procedure for the evaluation of vehicle dynamics and driver control performance. The slalom driving task could simulate the driver's steering behaviours when they need to make a lane change or over-taking manoeuvre, which is also a common driving response to a take-over request given by the partially automated driving vehicle [25]. The most critical intention prediction is made when the driver first take-over the vehicle to generate an advanced prediction for the steering direction and steering performance. Hence, the Slalom driving task is suitable to measure driver steering behaviors in this study.

The participants were required to use single-hand driving mode in the first with hands on the 3-clock positions. Then, they should perform four to six seconds of steering by following a constant period angle signal. The target sine-wave-like signal has about 60-degree amplitude with 0.25 Hz frequency to mimic the real-world steering behaviour. The driving cycle will be repeated three to five times for each participant, respectively. In general, each participant will be measured with a 12 to 15 period of steering with around five seconds steady holding phase after every three to five continuous steering. The main objective of this task is to simulate the real-world steering manoeuvre and gain the naturalistic neuromuscular dynamics from the participants. The data that support all the findings of this study are available from the corresponding authors upon reasonable research-purpose request.

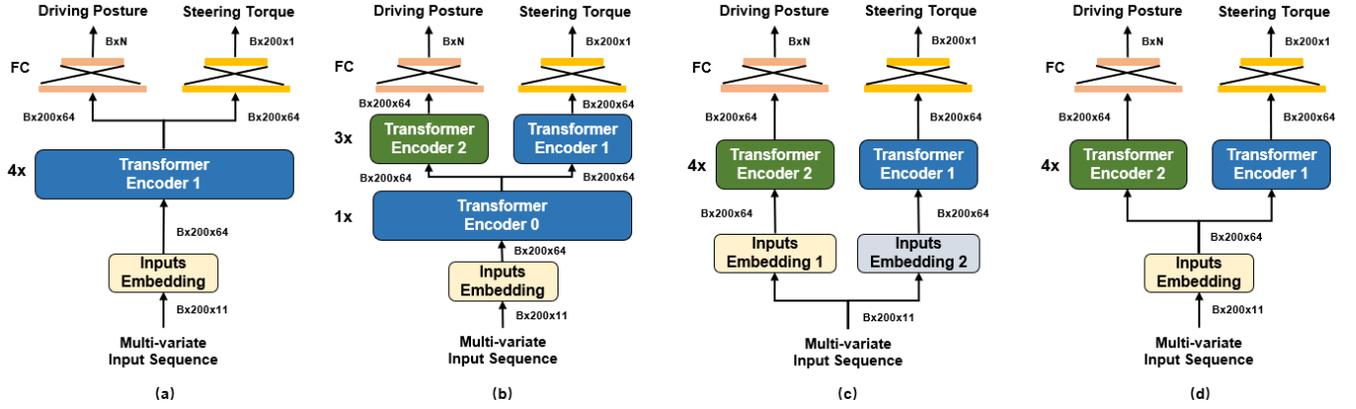

Figure 3. The proposed multi-task time-series transformer networks. (a) uses a shared embedding layer and three shared transformer encoder layers for the two tasks. (b) uses a shared embedding layer, one shared transformer encoder layer, and two personalized transformer encoder layers for the two tasks. (c) uses different embedding and encoder layers for the two tasks. (d) has similar architecture with (b) but get rid of the shared transformaer encoder layer.

### B. EMG Signal Processing

The EMG electrodes are placed at the proper position for the participants in the first. Then, the participants can take at least three minutes to prepare themselves on the testbed and sit in a preferred position to make sure they are relaxed. When the experiment starts, the EMG signals are collected and checked. The baseline noise is recorded and filtered for each channel. The steering torque for each participant is ranged from $-5N \cdot m$ to $5N \cdot m$, and the EMG signals are all ranged from -5 mv to 5 mv. A band-pass filter between 10 Hz to 500 Hz is then applied to the recorded EMG signals to remove outliers and interference. The time-series EMG data is divided into sequences with a fixed length of 200. A 200 ms prediction horizon without overlap is used as the sliding window for sequential data generation. The reasons for choosing the 200 ms horizon will be explained in the next section. The historical EMG and the steering torque signals will be used as the model input, and the future sequential steering torque is the target output of the model. All the signals are normalized to -1 to 1 before feeding into the model to avoid the adverse impact of a different unit.

The input ($X$) and target outputs $Y \epsilon [Y_t, Y_c]$ of the model can be denoted as:

$$X = [E_1, E_2, \cdots, E_9, E_{10}, Y_h] \quad (1)$$
$$E_i = [e_t, \cdots e_{t-200}] \quad (2)$$
$$Y_t = [st_{t+1}, \cdots, st_{t+200}] \quad (3)$$

where $X$ is the model input at each time step, $E_i$ represents the i[th] EMG feature sequence that consists of the last 200 ms data, $Y_t$ is the predicted torque, which contains the future 200 ms torque information, $Y_h$ is the history steering torque information in the past 200 ms, and $st_{t+1}$ is the steering torque at time $t + 1$. Hence, the dimension of the model input sequence ($X$) is $11 \times 200$, and the dimension of the output sequence ($Y_t$) is $1 \times 200$. $Y_c$ is the predicted driving postures, which can be one of the three categories for the single-hand driving mode.

### IV. MTS-TRANS MODEL DESIGN

In this section, the model development and implementation of the MTS-Trans models are introduced.

### A. MTS-Trans Architecture

In this study, four different MTS-Trans architecture is developed. Based on (1), a multi-variate input sequence $X \in \mathbb{R}^{l \times d}$ of length is fed $l$ and dimension $d$ into the sequential transformer networks. In general, all four architectures have similar blocks, including 1) input embedding, 2) transformer encoders, and 3) fully connected prediction heads.

**Embedding layer:** the original embedding layer for discrete tokens in [20] is replaced with the fully connected layer. The fully connected layer only applied to the feature dimension $d$ and transform the original 11 dimensions into 64 dimensions. The embedding layer is the only layer applied before the transformer encoder and no positional encoding is injected into the time-series data. The main reason is we found simply adding $cosine$ and $sine$ features as used in [20] into the input does not guarantee to give a better performance, however, adding positional encoding can be very likely to decrease the prediction accuracy in this study.

**Transformer encoder:** the transformer network is used in this part to capture both the long-term and short-term dependencies in the time-series input based on the multi-head self-attention mechanism. The transformer network avoids sequential modelling and prediction like LSTM, but treat the past sentence as a whole and assign self-attention along with the sequences. The transformer is briefly introduced here and can refer to [20] for detailed information. The multi-head self-attention mechanism in the encoder output $H \in \mathbb{R}^{l \times d}$ of the same size as $X$ by attending over given $l$ key-value pairs $K \in \mathbb{R}^{l \times d}, V \in \mathbb{R}^{l \times d}$:

$$H = Attention(Q, K, V) = softmax\left(\frac{QK^T}{\sqrt{m}}\right)V \quad (4)$$

Where $Q, K, V$ are query, key, and value matrix for the self-attention module. The scaled-dot product attention in (4) allows compressing the matrix $V$ into smaller representative bedding for simplified inference in the downstream neural network operations [26]. The scaled-dot product attention scales the weights by the $\sqrt{m}$ term to reduce the variance of the weights, facilitating stable training. Then the standard multi-head attention mechanism is applied to concatenate several self-attention heads. In this study, four transformer encoder layers are used in the four different architectures and each layer has eight self-attention heads.

**Fully-connected prediction heads:** on the top of the transformer encoder layers is the FC prediction heads. The FC prediction heads follow the following architectures:

$$O_{dp} = fc_{2\_dp}(Dropout(Relu(fc_{1\_dp}(H_{trans})))) \quad (5)$$

$$O_{st} = fc_{2\_st}(Dropout(Relu(fc_{1\_st}(H_{trans})))) \quad (6)$$

where $O_{dp} \in \mathbb{R}^{N_d \times 1}$ and $O_{st} \in \mathbb{R}^{l \times 1}$ are the predicted output for the driving postures and steering torque, $N_d$ is the number of driving postures. The output state $H_{trans}$ from the transformer, the layer is flattened into a one-dimension vector and passed into two FC layers. The output size of $fc_{1\_dp}$ and $fc_{1\_st}$ is 1024 uniformly.

### B. Loss Function

The optimization of the MTS-Trans model follows a multi-task learning framework. We jointly optimize the model on the two tasks and train the model in an end-to-end fashion. The overall training loss $L_T$ is a combination of the two individual losses, where the Cross-Entropy loss is used for the posture recognition task and the mean absolute error is used for the steering torque prediction. To deal with the homoscedastic uncertainty during model learning, the weighted loss function is used [27]. The overall loss function for MTS-Trans is denoted as follows.

$$L_T = W_1 L_p + W_2 L_{st} + \sum_{i=1}^{nT} \log \sigma_i \quad (7)$$

where $W_i = \exp(-\log \sigma_i^2)$ is the trainable weight for each sub-loss term considering the homoscedastic uncertainty or the observation noise $\sigma_i$ for the specific task ($\sigma_i$ initialized to zero), $nT$ equals to two, which is the number of tasks, and $L_p$, and $L_{st}$ are the loss of the posture recognition and steering torque prediction, respectively.

### C. Model Implementation

In this study, 9894 sequences are generated from the 3-clock driving posture scenarios, 9200 sequences are from the 130-clock categories, and 7660 sequences are collected from the 12-clock scenarios. 80% of the sequence is used for model training and the rest for model testing. The Adam optimizer is used for model optimization [28]. An initial learning rate (LR) of $1e-3$ is used and a step LR schedule is used to decay the LR by 0.1 every 100 epochs. The batch size is 64 and the maximum epoch is 300. The model is developed using PyTorch and is trained on a single NVIDIA Tesla P100 GPU. In general, the training process takes around three hours to finish.

## V. EXPERIMENT RESULTS

In this section, the model is evaluated based on several evaluation metrics and is compared with multiple baseline methods. Then, the different impacts of driving postures and driving modes on the steering intent prediction are quantitatively analysed.

### A. Baselines and Evaluation Metrics

To evaluate the continuous torque prediction results, two evaluation metrics, namely, the root-mean-square error (RMSE) and balanced RMSE (BRMSE) are used. To evaluate the driving posture recognition performance, another two-evaluation metrics, namely, the accuracy and F1 scores are used. The RMSE is calculated as follows.

$$RMSE = \frac{1}{N}\sum_{j=1,2,\cdots N} \sqrt{\frac{1}{L}\sum_{i=1,2,\cdots L}\left((\hat{x}_{ji} - x_{ji})^2\right)} \quad (8)$$

where $N$ is the number of sequences used for model testing, $L$ is the length of each sequence, which is 200 according to the time delay analysis. $\hat{x}_{ji}$ is the ith prediction in the sequence $j$, and $x_{ji}$ is the ground truth value.

To avoid the adverse impact of the data unbalanced property on the model evaluation results, the BRMSE metric split the data into bins and calculated the mean RMSE for each bin, and the final average BRMSE is then calculated [29]. As the steering torque is mostly ranged between $-5 \, N \cdot m$ to $5 \, N \cdot m$, ten bins are selected with each bin size having a $1 \, N \cdot m$ scale. The BRMS is represented as follows:

$$BRMSE_{d,k,j} = \frac{1}{N}\sum_i RMSE_i \quad (9)$$

$$BRMSE = \frac{1}{N_d}\sum BRMSE_{d.k.j} \quad (10)$$

where $BRMSE_{d,k,j}$ is the BRMSE for the $j^{th}$ bin, $d \coloneqq 1 \, N \cdot m$ is the bin size, $k \coloneqq 10 \, N \cdot m$ is the maximum range of steering torque, $RMSE_i$ is the RMSE for $i^{th}$ sequence in bin $j$, $N$ is the number of sequences in $j^{th}$ bin, $N_d \coloneqq 10$ is the overall number of bins, and $BRMSE$ is the final calculated balanced RMSE.

The accuracy and F1 score for the driving postures are defined as:

$$Acc = N_c/N_t \quad (11)$$
$$F1 = Tp/(Tp + (FP + FN)/2) \quad (12)$$

where $N_c$ is the number of correct classifications, and $N_t$ is the total number of samples, $Tp$, $FP$, and $FN$ are the number of true positives, false positives, and false negatives.

To evaluate the model performance on the different tasks, several baseline methods are introduced in this study.

1. **Random Prediction (ZP).** Always predict zero torque value which is simply used as a baseline. For posture recognition, ZP randomly guesses one posture.
2. **Feedforward Neural Network (FFNN).** An FFNN model with similar architecture to the prediction head of the MTS-Trans models is shown in Figure 3.

3. **LSTM Models (LSTM/Bi-LSTM).** A two-layer LSTM model (with or without bi-directional connections) with 64 feature dimensions in the hidden state.
4. **GRU Models (GRU/Bi-LSTM).** Similarly, a two-layer GRU model (with or without bi-directional connections) with 64 feature dimensions in the hidden state.
5. **Single task transformer.** Similar architecture with the MTS-Trans models as shown in Figure 3 but only have one branch and one prediction head each time for single-task learning.
6. **MTS-Trans**. Multi-task sequential transformer models with different fusion methods as shown in Figure 3.

*B. Single-Hand Driving Results*

In this part, the single-hand driving mode is analysed. The experiment results for single-hand driving mode is shown in Table 2. It can be concluded from Table 2 that the proposed MTS-Trans model can efficiently predict the future steering torques as well as recognize the driving postures. Specifically, the MTS-Trans models achieved around 90% accuracy in the posture recognition task. Meanwhile, accurate prediction of future steering torque is achieved with 0.0678 $N \cdot m$ RMSE with the MTS-Trans3 model. It is also shown that the third MTS-Trans model, which uses separate embedding layers and transformer encoder layers generated the most accurate results compared to the other methods.

When comparing multi-task learning with the single-task learning method, introducing extra tasks can improve the overall performance. Specifically, when simultaneously learning the postures, the RMSE of the steering torque prediction decreased to 0.0564 $N \cdot m$ from 0.0676 $N \cdot m$. Similarly, the posture recognition accuracy when considering torque prediction increased to 90.16% from 87.74%. Therefore, it is interesting to show that simultaneously learning these two tasks and optimizing the model parameters can improve the overall performance in both steering torque prediction and driving posture recognition.

TABLE 2. EXPERIMENT RESULTS FOR THE SINGLE-HAND DRIVING MODE

| Methods | Torque Prediction | | Posture Recognition | |
|---|---|---|---|---|
| | RMSE | BRMSE | Acc [%] | F1 [%] |
| ZP | 2.4267 | 2.4266 | 33.33 | 33.33 |
| FFNN | 0.1743 | 0.2111 | 73.86 | 73.96 |
| GRU | 0.1022 | 0.1290 | 72.88 | 72.64 |
| LSTM | 0.0832 | 0.0896 | 84.04 | 84.07 |
| BiGRU | 0.0768 | 0.0944 | 84.92 | 84.91 |
| BiLSTM | 0.0778 | 0.0865 | 84.79 | 84.84 |
| SSTrans | 0.0794 | 0.0977 | 87.74 | 87.82 |
| MTSTrans1 | 0.0892 | 0.1102 | 89.73 | 89.73 |
| MTSTrans2 | 0.0678 | 0.0895 | 90.10 | 90.10 |
| MTSTrans3 | 0.0564 | 0.0676 | 90.10 | 90.16 |
| MTSTrans4 | 0.0737 | 0.0901 | 89.40 | 89.37 |

The confusion matrix generated with the MTS-Trans3 model is shown in Figure 4. It is shown that as a middle-state, 130-clock driving posture is more likely to be misclassified and for the other two postures, it is more likely to be misclassified into the 130-clock. Specifically, 117 samples from the 130-clock are misclassified into the 3-clock case, and 106 samples are misclassified into the 12-clock. For the 3-clock case, 156 samples are misclassified to the 130-clock samples and only very few cases are misclassified into the 12-

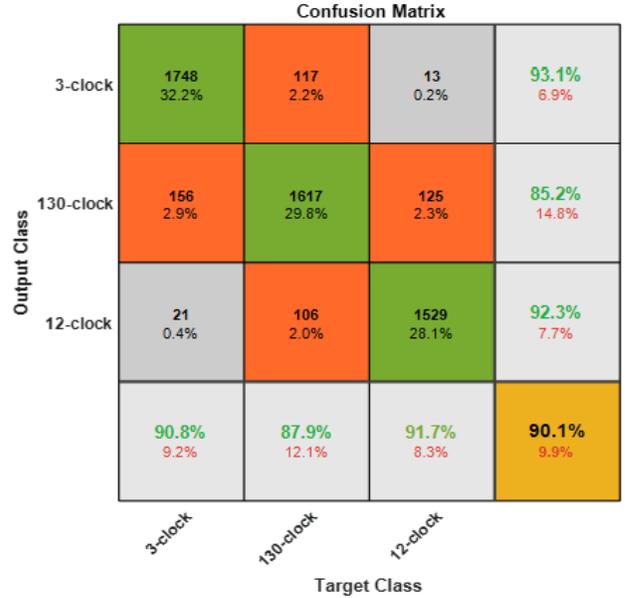

Figure 4. Confusion matrix for the single-hand driving mode driving posture recogniton with MTS-Trans3. The diagonal correspond to observations that are correctly classified while the off-diagonal cells correspond to incorrectly classified observations. The precision and recall are shown in the far right column and bottom row, respectively.

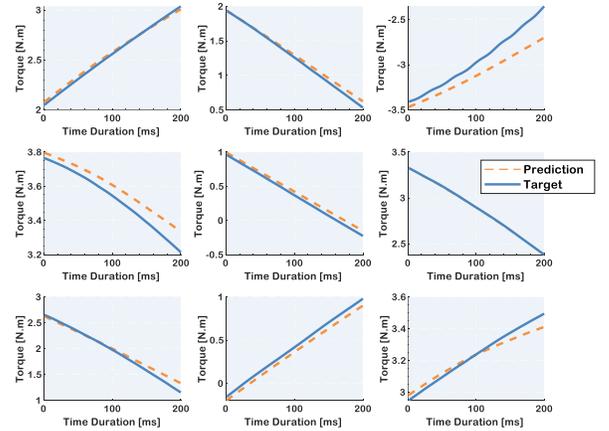

Figure 5. Steering torque prediction results from a single-hand driving scenario with MTS-Trans3 model.

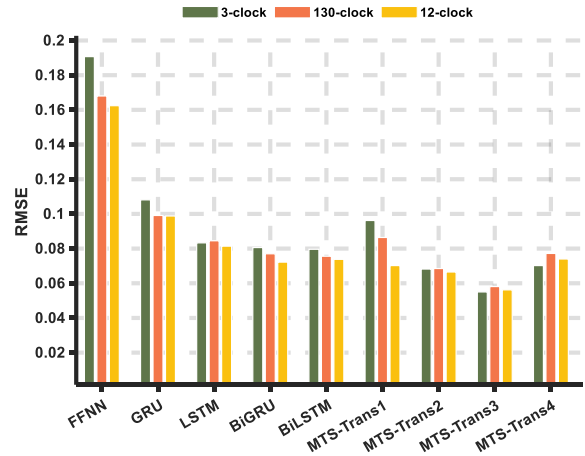

Figure 6. Comparison of the RMSE of steering torque prediction regarding different models and driving postures.

clock. Similar results can be found in the 12-clock case as well. Several examples of steering torque are shown in Figure 5. It is shown that the MTS-Trans model can make a reasonable prediction for future steering torques. The comparison of RMSE of the steering torque regarding different driving postures is shown in Figure 6. It can be found that the 12-clock driving postures usually lead to fewer prediction errors compared to the other two postures. While the 3-clock driving postures are more likely to give higher prediction errors.

## VI. CONCLUSION

This study develops a multi-task learning-based driving steering behaviour modelling framework using a sequential transformer network. The model can accurately estimate future steering torques and driving postures according to the neuromuscular dynamics. It is shown that a 200 ms prediction horizon can be used to predict the future steering torque based on historical EMG data. Then, the MTS-Trans model achieved a high precise prediction of future steering torques with a low RMSE error $0.0564\ N \cdot m$. It is also shown that the MTS-Trans model can also achieve around a 90% recognition rate for the driving postures based on the neuromuscular dynamics. As a middle stage, 130-clock driving postures are more likely to be misclassified into the other groups. The proposed network can realize precise sequential prediction for future steering intention, which will benefit future shared steering control design for the intelligent and automated driving vehicle towards a better human-vehicle-collaboration system.